\documentclass[a4paper,11pt]{article}
\pdfoutput=1
\usepackage[a4paper,left=2.73cm,right=2.7cm,top=3cm,bottom=3.5cm]{geometry}

\usepackage{amsfonts,amsmath,amssymb}
\usepackage{graphicx,caption,subcaption}
\usepackage[usenames, dvipsnames]{color}

\usepackage[colorlinks=true,linktocpage=true,linkcolor=blue,citecolor=blue]{hyperref}

\addtocontents{toc}{\protect\setcounter{tocdepth}{2}}
\numberwithin{equation}{section}

\def\cI{{\cal I}}

\def\cR{{\cal R}}

\newcommand{\be}{\begin{equation}}
\newcommand{\ee}{\end{equation}}

\begin{document}

\begin{titlepage}

\thispagestyle{empty}

\begin{center}

{\LARGE \textbf{\mbox{BPS black hole horizons from massive IIA}}   }

\vspace{40pt}
		
{\large \bf Adolfo Guarino}
		
\vspace{25pt}
		
{\normalsize  
Universit\'e Libre de Bruxelles (ULB) and International Solvay Institutes,\\
Service  de Physique Th\'eorique et Math\'ematique, \\
Campus de la Plaine, CP 231, B-1050, Brussels, Belgium.}

\vspace{20pt}

\texttt{aguarino@ulb.ac.be}

\vspace{20pt}
				
\abstract{
\noindent The maximal four-dimensional supergravity with a dyonic ISO(7) gauging that arises from the reduction of massive IIA on a six-sphere has recently been shown to accommodate static BPS black holes with hyperbolic horizons. When restricted to the $\mathcal{N}=2$ subsector that retains one vector multiplet and the universal hypermultiplet, the attractor mechanism was shown to fix both the vector charges and the scalar fields at the horizon to a unique configuration in terms of the gauging parameters. In order to assess the (non-)uniqueness of BPS black hole horizons from massive IIA, we extend the study of the attractor mechanism to other $\mathcal{N}=2$ subsectors including additional matter multiplets. We note that, while extending the hypermultiplet sector does not modify the set of solutions to the attractor equations, the inclusion of additional vector multiplets results in new hyperbolic/spherical horizon configurations containing free parameters. The model with three vector multiplets and the universal hypermultiplet, which is the massive IIA analogue of the STU-model from M-theory, may play a relevant role in massive IIA holography.}

\end{center}

\end{titlepage}

\tableofcontents

\hrulefill
\vspace{10pt}

\section{Motivation and outlook}
\label{sec:intro}

The study and characterisation of supersymmetric and asymptotically AdS$_{4}$ black holes in four-dimensional $\,\mathcal{N}=2\,$ gauged supergravity \cite{Caldarelli:1998hg,Cacciatori:2009iz,DallAgata:2010ejj,Hristov:2010ri} has recently received new attention in light of the gravity/gauge correspondence. Of special interest are the $\,\mathcal{N}=2\,$ gauged supergravities with a known embedding in string/M-theory \cite{Halmagyi:2013sla,Halmagyi:2014qza} due to the holographic description of their asymptotically AdS$_{4}$ black hole solutions in terms of RG-flows across dimensions \cite{Cucu:2003yk}. Especially striking results have been obtained in the context of BPS black holes from M-theory, providing non-trivial precision tests of the gravity/gauge correspondence beyond anti-de Sitter backgrounds \cite{Benini:2015eyy,Benini:2016rke}.

Eleven-dimensional supergravity, the low-energy limit of M-theory, can be consistently reduced on a seven-sphere to a maximal SO(8)-gauged supergravity in four dimensions \cite{deWit:1982ig,deWit:1986iy}. Within this theory, the so-called \mbox{STU-model} has played a central role for black holes. The model describes the $\,\textrm{U}(1)^4\,$ invariant subsector of the SO(8)-gauged supergravity \cite{Duff:1999gh,Cvetic:1999xp} which is an $\,\mathcal{N}=2\,$ gauged supergravity coupled to three vector multiplets in presence of $\,\textrm{U}(1)\,$ Fayet--Iliopoulos (FI) gaugings where the four FI parameters are identified. The black hole solutions studied in \cite{Cacciatori:2009iz,Benini:2015eyy} describe BPS flows interpolating between a maximally supersymmetric AdS$_{4}$ solution in the ultraviolet (UV) dual to ABJM \cite{Aharony:2008ug}, the superconformal field theory on a stack of \mbox{M2-branes}, and an $\,\textrm{AdS}_{2} \times \Sigma_{2}\,$ geometry in the near-horizon region. The classification of horizon configurations  can be performed by virtue of the attractor mechanism \cite{Cacciatori:2009iz,DallAgata:2010ejj,Hristov:2010ri,Halmagyi:2013qoa,Benini:2015eyy} which fixes the values of the scalars at the horizon in terms of the vector charges. At leading order, the gravitational entropy density associated with the horizons is obtained in terms of the charges using the Bekenstein--Hawking formula \cite{Bekenstein:1973ur,Hawking:1974sw}. Such a gravitational entropy density has been shown to nicely match the expression for the topologically twisted index computed in the (large $N$) dual field theory \cite{Benini:2015eyy,Benini:2016rke} (see also \cite{Cabo-Bizet:2017jsl}).

Upon compactification on a circle, eleven-dimensional supergravity reduces to massless IIA supergravity in ten dimensions. However, unlike the former, the latter is known to admit a deformation in terms of a mass parameter $\,m\,$ \cite{Romans:1985tz}. When $\,m \neq 0\,$ the connection to eleven-dimensional supergravity is lost rendering the massive IIA supergravity an independent theory.\footnote{See \cite{Gaiotto:2009mv,Gaiotto:2009yz} for holographic aspects of massive IIA on $\mathbb{CP}^3$ and deformations of the ABJM theory.} Similarly to the eleven-dimensional theory, massive IIA supergravity can be consistently reduced on a six-sphere to a maximal ISO(7)-gauged supergravity in four dimensions \cite{Guarino:2015vca} of the class investigated in \cite{Dall'Agata:2014ita}. In this theory, various types of BPS black holes have recently been found within the SU(3) invariant subsector \cite{Guarino:2017eag}. This subsector describes $\,\mathcal{N}=2\,$ gauged supergravity coupled to one vector multiplet and the universal hypermultiplet\footnote{Supersymmetric solutions of four-dimensional $\,\mathcal{N}=2\,$ supergravity models with vector and hypermultiplet sectors have been studied in \cite{Hristov:2009uj,Hristov:2010eu,Halmagyi:2011xh,Chimento:2015rra}.}, and the gauging, specified by the gauge coupling $\,g\,$ and the mass parameter $\,m$, is identified with a group $\,\textrm{G}=\mathbb{R} \times \textrm{U}(1)_{\mathbb{U}}\,$ of abelian isometries of the hypermultiplet moduli space \cite{Guarino:2015qaa}. The presence of the universal hypermultiplet in four dimensions is mandatory in order to accommodate the non-trivial ten-dimensional dilaton upon reduction on the six-sphere. The black hole solutions found in \cite{Guarino:2017eag} describe BPS flows interpolating between an $\,{\textrm{AdS}_{2} \times \textrm{H}^{2}}\,$ geometry in the near-horizon region and various UV asymptotic behaviours: charged $\textrm{AdS}_{4}$, non-relativistic scaling behaviours and the domain-wall DW$_{4}$ (four-dimensional) description of the D2-brane in massive IIA. The horizon configurations, namely the scalar fields and the vector charges at the horizon, turned out to be uniquely specified in terms of the gauging parameters $\,(g,m)\,$ related to the inverse radius of the six-sphere and the Romans mass parameter, modulo a $\mathbb{Z}_{2}$ reflection of the charges.

In this note we make some progress in the classification of BPS black hole horizon configurations in $\,\mathcal{N}=2\,$ supergravity models that arise from the reduction of massive IIA on the six-sphere. To this end, we extend the canonical setup with one vector multiplet and the universal hypermultiplet studied in \cite{Guarino:2017eag} by adding extra matter multiplets. Two cases are investigated:
\begin{itemize}
\item[$i)$] One vector multiplet and two hypermultiplets in the image of a c-map.
\item[$ii)$] Three vector multiplets and the universal hypermultiplet.
\end{itemize}
In the former case the attractor equations force the scalars in the extra hypermultiplet to vanish at the horizon, thus reducing this case to the one investigated in \cite{Guarino:2017eag}. In the latter case the extension of the vector sector proves more interesting.  The massive IIA model with three vector multiplets and the universal hypermultiplet is the analogue of the STU-model from \mbox{M-theory}, although there are some fundamental differences. For instance, while the STU-model from M-theory has a maximally symmetric AdS$_{4}$ vacuum dual to the superconformal ABJM theory on the M2-brane, the (massive) IIA counterpart is a DW$_{4}$ solution with a non-trivial profile for the dilaton in the universal hypermultiplet reflecting the non-conformality of the dual SYM(-CS) theory on the D2-brane \cite{Guarino:2016ynd}. The interplay between the scalars in the vector multiplets and the non-trivial dilaton in the universal hypermultiplet complicates the analysis of BPS flows. However, as we show in this note, the attractor equations governing the BPS horizon configurations can still be solved in full generality. The resulting horizon configurations are specified in terms of the gauging parameters $\,(g,m)\,$ together with four continuous parameters, and can have hyperbolic or spherical symmetry. The gravitational entropy density associated with these horizons may play an important role in precision tests of the massive~IIA on S$^{6}$/SYM-CS duality \cite{Guarino:2015jca,Schwarz:2004yj} beyond anti-de Sitter backgrounds.

\section{$\mathcal{N}=2\,$ gaugings and attractor mechanism from massive IIA}
\label{sec:N=2&BH}

Massive IIA ten-dimensional supergravity can be consistently reduced on S$^6$ down to a four-dimensional maximal supergravity with a dyonic ISO(7) gauging \cite{Guarino:2015vca}. This maximal supergravity admits various further truncations to $\,\mathcal{N}=2\,$ subsectors characterised by a compact subgroup $\,\textrm{G}_{0} \subset \textrm{ISO}(7)\,$ under which the fields retained in the truncation do not transform (singlets). The subsector with $\textrm{G}_{0}=\textrm{SU}(3)$ invariance has proved very successful in the study of AdS$_{4}$ \cite{Guarino:2015qaa}, domain-wall \cite{Guarino:2016ynd} and black hole \cite{Guarino:2017eag} solutions that can systematically be uplifted to ten dimensions\footnote{See \cite{Pang:2015vna} for the uplift of the AdS$_{4}$ vacuum preserving $\,\mathcal{N}=3\,$ supersymmetry and $\textrm{G}_{0} = \textrm{SO}(4)$ \mbox{found in \cite{Gallerati:2014xra}}.} by using the uplifting formulas of \cite{Guarino:2015vca} (see \textit{e.g.} \cite{Guarino:2015jca,Varela:2015uca}). However there are other $\,\mathcal{N}=2\,$ truncations based on different subgroups $\,\textrm{G}_{0}\,$ that are yet to be explored. This is what we set up to do in this note.

\subsection{Abelian hypermultiplet gaugings}
\label{sec:N=2}

With the aim of gaining new insights into the general structure of BPS black hole horizon configurations from massive IIA, we investigate various $\,\mathcal{N}=2\,$ truncations of the dyonically-gauged ISO(7) supergravity. They have an abelian gauge group $\,{\textrm{G}=\mathbb{R} \times \textrm{U}(1)_{\mathbb{U}}} \,$ and are described by a Lagrangian of the form
%
\begin{equation}
\label{Lagrangian_N2}
\begin{array}{lll}
L_{\textrm{mIIA}} &=&   \left( \frac{R}{2} - V \right) *  1 - K_{i \bar{j}} \,  dz^{i} \wedge * \, d\bar{z}^{\bar{j}} - h_{uv} \, Dq^{u} \wedge * \, Dq^{v}   \\[2mm]
&+& \frac{1}{2} \, \cI_{\Lambda \Sigma} \,  \mathcal{H}^\Lambda \wedge * \, \mathcal{H}^\Sigma +  \frac{1}{2} \, \cR_{\Lambda \Sigma} \, \mathcal{H}^\Lambda \wedge \mathcal{H}^\Sigma \\[2mm]
&-& \frac{1}{2} m\,  \mathcal{B}^{0} \wedge d \tilde{\mathcal{A}}_0 - \frac{1}{8} \, g \,  m\,  \mathcal{B}^{0} \wedge \mathcal{B}^{0} \ .
\end{array}
\end{equation}
The (dynamical) field content of the various models studied in this note consists of the supergravity multiplet coupled to $\,n_{v}\,$ vector multiplets and $\,n_{h}\,$ hypermultiplets.

The complex scalars $\,z^{i}\,$ in the vector multiplets, with $\,i=1,...,n_{v}\,$, serve as coordinates in a special K\"ahler (SK) manifold $\,\mathcal{M}_{\textrm{SK}}\,$. As the gauging is abelian, they must be neutral
\begin{equation}
\label{Dz}
D z^{i}=d z^{i} \ .
\end{equation}
In order to describe the dynamics of the vector multiplets, namely the kinetic terms for scalars and vectors as well as the generalised theta angles in (\ref{Lagrangian_N2}), we adopt the same conventions as in \cite{Klemm:2016wng,Guarino:2017eag}. Introducing a symplectic product of the form 
\begin{equation}
\left\langle U , V \right\rangle \equiv  U^{M} \Omega_{MN}V^{N} = U_{\Lambda} V^{\Lambda} - U^{\Lambda} V_{\Lambda} \ ,
\end{equation}
where $\,\Omega_{MN}\,$ is the (antisymmetric) invariant matrix of $\,\textrm{Sp}(2 \, n_{v}+2)\,$ and $\,\Lambda=0,...,n_{v}\,$, the kinetic terms for the scalars $\,z^{i}\,$ are determined by a K\"ahler potential
\begin{equation}
\label{K_pot}
K=-\log(i \left\langle X, \bar{X} \right\rangle) \ .
\end{equation}
This is in turn expressed in terms of holomorphic sections $\,X^{M}(z^{i})=(X^{\Lambda},F_{\Lambda})\,$ satisfying $\,F_{\Lambda} = \partial \mathcal{F}/\partial X^{\Lambda}\,$ for a prepotential $\,\mathcal{F}(X^{\Lambda})\,$ that is a homogeneous function of degree two. In terms of the K\"ahler potential (\ref{K_pot}), the metric in $\,\mathcal{M}_{\textrm{SK}}\,$ is given by
\begin{equation}
\label{dsSK_N2}
ds_{\textrm{SK}}^2 = K_{i \bar{j}} \, dz^{i} \,  d\bar{z}^{\bar{j}}  
\hspace{8mm} \textrm{ with } \hspace{8mm}
K_{i \bar{j}} = \partial_{z^{i}} \partial_{\bar{z}^{\bar{j}}} K \ .
\end{equation}
The kinetic terms and generalised theta angles for the (dynamical) vectors $\,\mathcal{A}^{\Lambda}\,$ are encoded in the matrix
\begin{equation}
\label{N_matrixN2}
\mathcal{N}_{\Lambda \Sigma} = \bar{F}_{\Lambda \Sigma}+ 2 \, i \,  \frac{\textrm{Im}(F_{\Lambda \Gamma})X^{\Gamma}\,\,\textrm{Im}(F_{\Sigma \Delta})X^{\Delta}}{\textrm{Im}(F_{\Omega \Phi})X^{\Omega}X^{\Phi}}
\hspace{5mm} \textrm{ where } \hspace{5mm}
F_{\Lambda \Sigma}=\partial_{\Lambda}\partial_{\Sigma} \mathcal{F} \ .
\end{equation}
More concretely, the relevant functions entering (\ref{Lagrangian_N2}) are obtained as $\,\mathcal{R}_{\Lambda \Sigma}\equiv \textrm{Re}(\mathcal{N}_{\Lambda \Sigma})\,$ and $\,\mathcal{I}_{\Lambda \Sigma}\equiv \textrm{Im}(\mathcal{N}_{\Lambda \Sigma})\,$, and can be used to define a symmetric, real and negative-definite scalar matrix
\begin{equation}
\label{M_scalar_matrix}
\mathcal{M}(z^{i}) = \left( 
\begin{array}{cc}
\mathcal{I} + \mathcal{R} \mathcal{I}^{-1} \mathcal{R}   & -\mathcal{R} \mathcal{I}^{-1} \\
- \mathcal{I}^{-1} \mathcal{R} & \mathcal{I}^{-1}
\end{array}
\right) \ .
\end{equation}

The real scalars $\,q^{u}\,$ in the hypermultiplets, with $\,u=1,...,4 \, n_{h}\,$, parameterise a quaternionic K\"ahler (QK) manifold $\,\mathcal{M}_{\textrm{QK}}\,$ with metric
\begin{equation}
ds_{\textrm{QK}}^2 = h_{uv} \, dq^{u} dq^{v} \ .
\end{equation}
In this work we focus on QK manifolds that lie in the image of a c-map \cite{deWit:1990na,deWit:1992wf,deWit:1993rr}. The metric in this class of QK manifolds takes the form
\begin{equation}
\label{dsQK_c-map}
\begin{array}{lll}
ds_{\textrm{QK}}^2 &=&  \widetilde{K}_{a \bar{b}} \, d\tilde{z}^{a} \, d\bar{\tilde{z}}^{\bar{b}}
+  \, d \phi \,  d \phi - \frac{1}{4} \, e^{2\phi} \,  ( d\vec{\zeta}\,)^{T}  \mathbb{C}\mathbb{M}_{4} \, d \vec{\zeta} \\[2mm]
	&+& \frac{1}{4} e^{4 \phi} \left[ d \sigma +\frac{1}{2}\, (\vec{\zeta}\,)^{T} \, \mathbb{C} \, d \vec{\zeta} \,  \right]  \,   \left[ d \sigma +\frac{1}{2}\, (\vec{\zeta}\,)^{T} \, \mathbb{C} \, d \vec{\zeta} \, \right]  \ ,
\end{array}
\end{equation}
with $\,\mathbb{C}=-\Omega\,$. The matrix $\,\mathbb{M}_{4}\,$ entering the first line in (\ref{dsQK_c-map}) depends on the complex scalars  $\,\tilde{z}^{a}\,$, with $\,a = 1, ..., n_{h}-1\,$, parameterising the special K\"ahler manifold $\,\mathcal{M}_{\widetilde{\textrm{SK}}}\,$ of the c-map. The remaining coordinates in $\,\mathcal{M}_{\textrm{QK}}\,$ form the set $\,\{\phi \,,\, \sigma \,,\, \zeta^{A} \,,\, \tilde{\zeta}_{A}\}\,$ with $\,A=0, ..., n_{h}-1\,$. We have also defined $\,\vec{\zeta}\equiv(\zeta^{A},\tilde{\zeta}_{A})\,$ in (\ref{dsQK_c-map}).

It is customary in $\,\mathcal{N}=2\,$ supergravity to arrange electric $\,\mathcal{A}^{\Lambda}\,$ and magnetic $\,\tilde{A}_{\Lambda}\,$ vectors into a symplectic vector  $\,\mathcal{A}^{M}=(\mathcal{A}^{\Lambda},\tilde{\mathcal{A}}_{\Lambda})\,$  with $\,M\,$ being a fundamental index of the electric-magnetic group $\,\textrm{Sp}(2 \, n_{v}+2)\,$. In the massive IIA models, both types of vectors participate in the gauging of the $\,{\textrm{G}=\mathbb{R} \times \textrm{U}(1)_{\mathbb{U}}} \,$ abelian isometries of $\,\mathcal{M}_{\textrm{QK}}\,$. As a result, the scalars $\,q^{u}\,$ in the hypermultiplets are charged under the gauging and minimally couple to the vector fields via covariant derivatives of the form
\begin{equation}
\label{Dq_N2}
Dq^{u} = \partial q^{u} - \mathcal{A}^{M} \, \Theta_{M}{}^{\alpha} \, k^{u}_{\alpha} \ .
\end{equation}
The Killing vectors $\,k_{\alpha}\,$ (with $\alpha=\mathbb{R}$ or $\mathbb{U}$) in (\ref{Dq_N2}) couple simultaneously to electric and magnetic vectors as dicated by a dyonic embedding tensor $\,\Theta_{M}{}^{\alpha}=(\Theta_{\Lambda}{}^{\alpha},\Theta^{\Lambda \, \alpha})\,$  with $\,{\Theta^{\Lambda \, \alpha} \neq 0}\,$. Consistency of the gauging requires an orthogonality constraint of the form $\,\left\langle   \Theta^{\alpha},\Theta^{\beta} \right\rangle=0\,$ \cite{deWit:2005ub}. This constraint is guaranteed for the dyonic embedding tensor underlying the massive IIA models, which takes the form
\begin{equation}
\label{Theta_tensor_N2}
\Theta_{M}{}^{\alpha} = \left(
\begin{array}{c}
\Theta_{\Lambda}{}^{\alpha} \\[2mm]
\hline\\[-2mm]
\Theta^{\Lambda \, \alpha} 
\end{array}\right)
= \left(
\begin{array}{cc}
\Theta_{0}{}^{\mathbb{R}} & \Theta_{0}{}^{\mathbb{U}} \\[2mm]
\Theta_{1}{}^{\mathbb{R}} & \Theta_{1}{}^{\mathbb{U}} \\[2mm]
\vdots & \vdots \\[2mm]
\Theta_{n_{v}}{}^{\mathbb{R}} & \Theta_{n_{v}}{}^{\mathbb{U}} \\[2mm]
\hline\\[-2mm]
\Theta^{0 \, \mathbb{R}} & \Theta^{0 \, \mathbb{U}} \\[2mm]
\Theta^{1 \, \mathbb{R}} & \Theta^{1 \, \mathbb{U}} \\[2mm]
\vdots & \vdots \\[2mm]
\Theta^{n_{v} \, \mathbb{R}} & \Theta^{n_{v} \, \mathbb{U}} \\[2mm]
\end{array}\right)
= \left(
\begin{array}{cc}
g & 0 \\[2mm]
0 & g \\[2mm]
\vdots & \vdots \\[2mm]
0 & g \\[2mm]
\hline\\[-2mm]
-m & 0 \\[2mm]
0 & 0 \\[2mm]
\vdots & \vdots \\[2mm]
0 & 0 
\end{array}\right) \ .
\end{equation}
The dyonic nature of the four-dimensional gauging has its origin in the Romans mass parameter $\,m\,$ of the ten-dimensional massive IIA supergravity \cite{Guarino:2015jca}. More specifically, it only affects the $\,\mathbb{R}\,$ factor of the gauge group which is associated with the isometry $\,k_{\mathbb{R}} =\partial_{\sigma}\,$ of the quaternionic manifold (\ref{dsQK_c-map}). This isometry is gauged by a linear combination of the graviphoton and its magnetic dual, as it can be seen from the covariant derivatives
\begin{equation}
\label{Dq_N2_massive_IIA}
Dq^{u} = \partial q^{u} -  (g \, \mathcal{A}^{0} - m \, \tilde{\mathcal{A}}_{0}) \,  k^{u}_{\mathbb{R}}  - g \,    \mathcal{A}_{\mathbb{U}}  \, k^{u}_{\mathbb{U}} \ .
\end{equation}
From (\ref{Dq_N2_massive_IIA}) one  also sees that $\,\sigma\,$ becomes a St\"uckelberg field. The $\,\textrm{U}(1)_{\mathbb{U}}\,$ factor of the gauge group associated with the isometry $\,k_{\mathbb{U}}\,$ is spanned by the electric vector $\,\mathcal{A}_{\mathbb{U}} \equiv \sum_{i} \, \mathcal{A}^{i}\,$.

The Romans mass also induces the topological term in the last line of (\ref{Lagrangian_N2}) which involves the magnetic graviphoton $\,\tilde{\mathcal{A}}_{0}\,$ and an auxiliary two-form tensor field $\,\mathcal{B}^{0}\,$. The presence of non-dynamical tensor fields in four-dimensional gauged supergravities with a dyonic gauging is a well understood phenomenon \cite{deWit:2005ub}. In addition to the topological term, the auxiliary tensor field $\,\mathcal{B}^{0}\,$ modifies the field strength of the electric graviphoton. Concretely, one has that
\begin{equation}
\label{Field-strengthsH}
\mathcal{H}^{0} = d \mathcal{A}^{0} + \tfrac{1}{2 } \, m \,  \mathcal{B}^{0}
\hspace{5mm} , \hspace{5mm} \mathcal{H}^{i} = d \mathcal{A}^{i} \ .
\end{equation}
The equation of motion for the magnetic graviphoton following from the Lagrangian (\ref{Lagrangian_N2}) and (\ref{Dq_N2_massive_IIA}) gives a duality relation between the auxiliary tensor field $\,\mathcal{B}^{0}\,$ and the covariant derivatives of the scalars $\,\sigma\,$ and $\,\vec{\zeta}\,$ in the hypermultiplet sector
\begin{equation}
\label{EOM_A0tilde}
\begin{split}
d \mathcal{B}^0  & = -  e^{4 \phi} * \! \left[ D \sigma +\frac{1}{2}\, (\vec{\zeta}\,)^{T} \, \mathbb{C} \, D \vec{\zeta} \,  \right] \ .
 \end{split}
\end{equation}
In addition the dual graviphoton is subject to a duality relation
\begin{equation}
\label{EOM_B0}
d \tilde{\mathcal{A}}_{0}+\tfrac{1}{2} \, g \, \mathcal{B}^{0}=\mathcal{I}_{0 \Sigma} * \mathcal{H}^\Sigma  + \mathcal{R}_{0 \Sigma} \, \mathcal{H}^\Sigma \ ,
\end{equation}
arising as the equation of motion for the tensor field $\,\mathcal{B}^0\,$.

The gauging of abelian isometries in $\,\mathcal{M}_{\textrm{QK}}\,$ induces a potential for the scalar fields in the vector multiplets and hypermultiplets. Using $\,\mathcal{N}=2\,$ symplectically covariant notation, it is given by \cite{Michelson:1996pn,deWit:2005ub}
\begin{equation}
\label{VN2}
V =  4 \,  \mathcal{V}^{M}  \, \bar{\mathcal{V}}^{N}    \, \mathcal{K}_{M}{}^{u}  \, h_{uv} \,  \mathcal{K}_{N}{}^{v}
+ \mathcal{P}^{x}_{M} \, \mathcal{P}^{x}_{N} \left( K^{i\bar{j}} \, D_{i}\mathcal{V}^{M} \, D_{\bar{j}} \bar{\mathcal{V}}^{N}  - 3 \, \mathcal{V}^{M} \, \bar{\mathcal{V}}^{N} \right) \ ,
\end{equation}
with $\,\mathcal{V}^{M} \equiv e^{K/2} \, X^{M}\,$ and $\,D_{i}\mathcal{V}^{M}=\partial_{z^{i}} \mathcal{V}^{M} + \frac{1}{2} (\partial_{z^{i}} K)\mathcal{V}^{M}\,$, and where we have introduced symplectic Killing vectors $\,\mathcal{K}_{M}  \equiv \Theta_{M}{}^{\alpha} \, k_{\alpha}\,$ and moment maps $\,\mathcal{P}_{M}^{x}  \equiv \Theta_{M}{}^{\alpha} \, P_{\alpha}^{x}\,$ in order to maintain symplectic covariance \cite{Klemm:2016wng}. Lastly, the Einstein-Hilbert term closes the description of the supergravity Lagrangian in (\ref{Lagrangian_N2}).

\subsection{Static BPS black holes and attractor mechanism}
\label{sec:attractor}

Static BPS black holes with spherical/hyperbolic symmetry have been extensively studied in the context of $\,\mathcal{N}=2\,$ gauged supergravity. Adopting the conventions of \cite{Klemm:2016wng}, the most general metric compatible with the symmetry takes the form
\begin{equation}
\label{ansatz_metric}
d s^2 = - e^{2 U} d t^2 + e^{-2 U} d r^2 + e^{2 (\psi - U)} \left( d \theta^2 + \left( \frac{ \sin \sqrt{\kappa} \, \theta }{ \sqrt{ \kappa } } \right)^2 \, d \phi^2 \right) \ ,
\end{equation}
with $\,\kappa=1\,$ (spherical horizon S$^2$) or $\,\kappa=-1\,$ (hyperbolic horizon H$^2$). The functions $\,e^{-2 U}\,$ and $\,e^{2 (\psi - U)}\,$ in the metric (\ref{ansatz_metric}) as well as the vectors $\,\mathcal{A}^{\Lambda}\,$ and $\,\tilde{\mathcal{A}}_{0}\,$, the tensor $\,\mathcal{B}^{0}\,$ and the scalars $\,z^{i}\,$ and $\,q^{u}\,$ are assumed to depend only on the radial coordinate $\,r\,$. 
The ansatz for the vectors takes the form
\begin{equation}
\label{ansatz_vectors}
\mathcal{A}^\Lambda = \mathcal{A}_t{}^\Lambda(r) \, d t - p^\Lambda \, \frac{ \cos \sqrt{\kappa} \, \theta }{ \kappa } \, d \phi 
\hspace{6mm} \textrm{ , } \hspace{6mm}
\tilde{\mathcal{A}}_0 = \tilde{\mathcal{A}}_{t \, 0}(r) \, d t - e_{0}  \, \frac{ \cos \sqrt{\kappa} \, \theta }{ \kappa } \, d \phi \ ,
\end{equation}
with constant magnetic $\,p^{\Lambda}\,$ and electric $\,e_{0}\,$ charges. The ansatz for the tensor field reads\footnote{The expression in (\ref{ansatz_tensor}) differs from the one in \cite{Klemm:2016wng} by a tensor gauge transformation \cite{Guarino:2017eag}.}
\begin{equation}
\label{ansatz_tensor}
\mathcal{B}^0 = b_0(r) \, \frac{ \sin \sqrt{\kappa} \, \theta }{ \sqrt{\kappa} } \, d \theta \wedge d \phi \ .
\end{equation}

As we will see in the next section, the scalars being charged under the vectors $\,\mathcal{A}^{i}\,$ are forced to vanish at the horizon by virtue of the attractor mechanism. Furthermore, they can be set to zero identically at the level of the Lagrangian without causing inconsistencies with the set of equations of motion derived from (\ref{Lagrangian_N2}). The latter restriction implies that there is no source term in the equations of motion for the vectors $\,\mathcal{A}^{i}\,$ which become of the form $\,{d(\mathcal{I}_{i \Sigma} * \mathcal{H}^\Sigma  + \mathcal{R}_{i \Sigma} \, \mathcal{H}^\Sigma)=0}\,$. For this reason, it is convenient to also introduce a set of auxiliary magnetic vectors
\begin{equation}
\label{ansatz_vector_dual}
\tilde{\mathcal{A}}_{i} = \tilde{\mathcal{A}}_{t \, {i}}(r) d t - e_{i} \, \frac{ \cos \sqrt{\kappa} \, \theta }{ \kappa } \, d \phi\ ,
\end{equation}
with constant electric charges $\,e_{i}\,$, which are subject to a set of duality relations of the form $\,{d\tilde{\mathcal{A}}_{i}=\mathcal{I}_{i \Sigma} * \mathcal{H}^\Sigma  + \mathcal{R}_{i \Sigma} \, \mathcal{H}^\Sigma}\,$. The analysis of equations of motion and duality relations is the straightforward generalisation of the one performed in \cite{Guarino:2017eag} for the model with one vector multiplet and the universal hypermultiplet and fits into the general analysis of \cite{Guarino:2015qaa}. 

In the near-horizon region the metric takes the form $\,\textrm{AdS}_{2} \times \Sigma_{2}\,$ with $\,\Sigma_{2}= \{\textrm{S}^{2} \,,\, \textrm{H}^{2}\}\,$. This sets the functions in (\ref{ansatz_metric}) to
\begin{equation}
\label{metric_horizon}
e^{2U} = \frac{r^2}{L^2_{\textrm{AdS}_{2}}}
\hspace{8mm} \textrm{ and } \hspace{8mm}
e^{2(\psi-U)} = L^{2}_{\Sigma_{2}} \ .
\end{equation}
On the other hand, BPS black holes are solutions to a set of first-order flow equations: they extremise a real function $\,2|W|\,$ defined in terms of a central charge $\,\mathcal{Z}\,$ and a superpotential~$\,\mathcal{L}\,$. This function, that must vanish at the horizon \cite{Klemm:2016wng}, is given by
\begin{equation}
\label{EqZ+L=0}
W=e^{U} (\mathcal{Z} + i \, \kappa \, L_{\Sigma_{2}}^2 \, \mathcal{L})= |W| \, e^{i \beta} \ ,
\end{equation}
and solves the Hamilton--Jacobi equation for the effective action obtained upon plugging the field ansatz (\ref{ansatz_metric})-(\ref{ansatz_vector_dual}) into the Lagrangian (\ref{Lagrangian_N2}). The central charge and the superpotential are obtained as
\begin{equation}
\mathcal{Z}(z^{i})=\left\langle \mathcal{Q} , \mathcal{V} \right\rangle
\hspace{10mm} \textrm{ and } \hspace{10mm}
\mathcal{L}(z^{i},q^{u})=\left\langle \mathcal{Q}^{x} \mathcal{P}^{x} , \mathcal{V} \right\rangle \ ,
\end{equation}
with $\,\mathcal{Q}^{x} \equiv \left\langle  \mathcal{P}^{x} , \mathcal{Q} \right\rangle\,$, and depend on a symplectic vector of charges
\begin{equation}
\label{Q_vector_attractor}
{\mathcal{Q}^M} = \left(  \,\, \mathfrak{p}^{0} \,\, , \,\,  p^{i}  \,\,,\,\,  \mathfrak{e}_{0}  \,\,,\,\, e_{i}  \,\, \right)^{T} \ ,
\end{equation}
where $\,\mathfrak{p}^{0}\equiv p^{0} + \tfrac{1}{2} \, m \, b_{0}(r)\,$ and $\, \mathfrak{e}_{0} \equiv e_0 + \tfrac{1}{2} \, g \, b_0(r)\,$. Assuming that the scalars approach the horizon with a constant value, \textit{i.e.} $\,{z^{i}}'={q^{u}}'=0\,$, the first-order flow equations become algebraic and determine the so-called attractor equations. The set of attractor equations for dyonic gaugings of $\,\mathcal{N}=2\,$ supergravity was derived in \cite{Klemm:2016wng}. It is given by\footnote{The charges $\,\mathcal{Q}\,$ in (\ref{attrac_eqs}) are understood as evaluated at the horizon, namely, $\,{\mathfrak{p}^{0}\equiv p^{0} + \tfrac{1}{2} \, m \, b^{h}_{0}}\,$ and $\, \mathfrak{e}_{0} \equiv e_0 + \tfrac{1}{2} \, g \, b^{h}_0\,$. From the set of first-order BPS equations it can be shown that $\,\mathcal{Q}'=0\,$ at the horizon \cite{Klemm:2016wng}.}
\begin{equation}
\label{attrac_eqs}
\begin{split}
\mathcal{Q} & =   \kappa \, L_{\Sigma_{2}}^{2} \, \Omega \, \mathcal{M} \, \mathcal{Q} ^{x} \, \mathcal{P}^{x} - 4 \, \textrm{Im}(\bar{\mathcal{Z}} \, \mathcal{V}) \ , \\
\dfrac{L_{\Sigma_{2}}^{2}}{L_{\textrm{AdS}_{2}}} & =  -2 \, \mathcal{Z} \, e^{-i \beta} \ , \\[2mm]
\left\langle  \mathcal{K}^{u} , \mathcal{V} \right\rangle & =  0 \ ,
\end{split}
\end{equation}
and must be supplemented with a charge quantisation condition
\begin{equation}
\label{quant_cond}
\mathcal{Q}^{x} \, \mathcal{Q}^{x} = 1 \ ,
\end{equation}
and a set of compatibility constraints of the form
\begin{equation}
\label{extra_constraints}
\mathcal{H} \, \Omega \, \mathcal{Q} = 0 
\hspace{8mm} \textrm{ and } \hspace{8mm}
\mathcal{H} \, \Omega \, \mathcal{A}_{t} = 0 \ ,
\end{equation}
where $\,\mathcal{H}=(\mathcal{K}^u)^{T} \, h_{uv} \, \mathcal{K}{}^{v}\,$. We refer the reader to the original work of \cite{Klemm:2016wng} for a detailed derivation of the attractor equations (\ref{attrac_eqs})-(\ref{extra_constraints}). 

The values of the scalars at the horizon configurations can be alternatively obtained by extremising an effective black hole potential. Due to the presence of the gauging in the hypermultiplet sector, such an effective potential takes the form \cite{Chimento:2015rra}
\begin{equation}
\label{Veff}
V_{\textrm{eff}} = \frac{\kappa-\sqrt{\kappa^2-4 \, V_{\textrm{BH}} \, V}}{2 \, V} \ ,
\end{equation}
with $\,V\,$ given in (\ref{VN2}) and where $\,V_{\textrm{BH}}=-\frac{1}{2} \mathcal{Q}^{\,T} \, \mathcal{M} \, \mathcal{Q}\,$ is the black hole potential in $\,\mathcal{N}=2\,$ ungauged supergravity \cite{Ferrara:1997tw} that depends on the charges and on the scalar-dependent matrix (\ref{M_scalar_matrix}). One then has that 
\begin{equation}
\label{Veff_extremisation}
\partial_{z^{i}}V_{\textrm{eff}} \big|_{z_{h}^{i},q_{h}^{u}} = 0
\hspace{5mm} , \hspace{5mm}
\partial_{q^{u}}V_{\textrm{eff}} \big|_{z_{h}^{i},q_{h}^{u}} = 0
\hspace{5mm} \textrm{ and }\hspace{5mm}
L^2_{\Sigma_{2}} = V_{\textrm{eff}}(z_{h}^{i},q_{h}^{u}) \ ,
\end{equation}
where we have denoted the values of the fields at the horizon with a subscript $_{h}\,$.
In the next section, we are solving the attractor equations (\ref{attrac_eqs})-(\ref{extra_constraints}) for various $\,\mathcal{N}=2\,$ supergravity models arising from the reduction of massive~IIA on the six-sphere.

\section{Models from massive IIA}

Here we investigate the existence of BPS horizon configurations with hyperbolic/spherical symmetry in various $\,\mathcal{N}=2\,$ truncations of the dyonically-gauged ISO(7) supergravity \cite{Guarino:2015qaa}. We start the section by reviewing the horizon configurations found in \cite{Guarino:2017eag} for the canonical model with one vector multiplet and the universal hypermultiplet, and then generalise the setup there by including additional matter multiplets.

\subsection{$\mathcal{M}_{\textrm{SK}}=\textrm{SU(1,1)}/\textrm{U}(1)\,$ and $\,\mathcal{M}_{\textrm{QK}}=\textrm{SU(2,1)}/(\textrm{SU}(2) \times \textrm{U}(1))\,$}
\label{sec:nv=1&nh=1}

Examples of four-dimensional static BPS black holes enjoying an embedding in massive IIA supergravity were presented in \cite{Guarino:2017eag}.\footnote{See also \cite{Kimura:2012yz} for AdS$_{4}$ black holes from massive IIA with vanishing electromagnetic charges.} The $\,\mathcal{N}=2\,$ model studied there corresponds to the truncation preserving $\,\textrm{G}_{0}=\textrm{SU}(3) \subset \textrm{ISO}(7)\,$ \cite{Guarino:2015qaa}. This model describes $\,\mathcal{N}=2\,$ supergravity coupled to a vector multiplet $\,(n_{v}=1)\,$ and the universal hypermultiplet $\,(n_{h}=1)\,$, and has
\begin{equation}
\mathcal{M}_{\textrm{SK}}=\frac{\textrm{SU(1,1)}}{\textrm{U}(1)}  
\hspace{8mm} \textrm{ and } \hspace{8mm}
\mathcal{M}_{\textrm{QK}}=\frac{\textrm{SU}(2,1)}{\textrm{SU}(2) \times \textrm{U}(1)} \ .
\end{equation}
The holomorphic sections used to describe the SK manifold are given by
\begin{equation}
\label{Xsections_nv=1&nh=1}
(X^{0}\,,\,X^{1}\,,\,F_{0}\,,\,F_{1}) =  (-z^3\,,\,-z\,,\,1\,,\,3z^2) \ , 
\end{equation}
which follow from a square-root prepotential
\begin{equation}
\label{F_prepot_nv=1&nh=1}
\mathcal{F} = - 2 \, \sqrt{X^{0}(X^{1})^3} \ .
\end{equation}

Denoting $\,z^{1}\equiv z = - \chi +  i e^{-\varphi}\,$ the scalar in the vector multiplet parameterising $\,\mathcal{M}_{\textrm{SK}}\,$ and $\,q^{u}=(\phi \, ,\, \sigma \,,\, \zeta^{0} \equiv \zeta \,,\, \tilde{\zeta}_{0} \equiv\tilde{\zeta})\,$ the four real scalars serving as coordinates in $\,\mathcal{M}_{\textrm{QK}}\,$, the scalar metrics entering (\ref{Lagrangian_N2}) take the form
\begin{equation}
\label{dsSK_nv=1&nh=1}
ds_{\textrm{SK}}^2  = \frac{3}{4} \frac{dz \, d\bar{z}}{(\textrm{Im} z)^2} \ ,
\end{equation}
and
\begin{equation}
\label{dsQK_nv=1&nh=1}
\begin{array}{lll}
ds_{\textrm{QK}}^2 &=&   d \phi \,  d \phi + \frac{1}{4} \, e^{2\phi} \,   \left( d \zeta \,  d\zeta + d \tilde{\zeta} \,  d \tilde{\zeta} \right) \\[2mm] 
&+& \frac{1}{4} e^{4 \phi} \left[ d \sigma + \tfrac{1}{2}  ( \tilde{\zeta} d \zeta - \zeta d \tilde{\zeta}  ) \right] \, \left[ d \sigma + \tfrac{1}{2}  ( \tilde{\zeta} d \zeta - \zeta d \tilde{\zeta}  ) \right]  \ .
\end{array}
\end{equation}
The kinetic terms and the generalised theta angles for the vectors are determined by the scalar-dependent matrix (\ref{N_matrixN2}) which reads
\begin{equation}
\label{NMat_nv=1&nh=1}
\mathcal{N}_{\Lambda \Sigma}  = 
\frac{1}{(2\, e^{\varphi } \, \chi +i )}
\left(
\begin{array}{cc}
 -\dfrac{e^{3 \varphi }}{(e^{\varphi } \, \chi -i )^2} & \dfrac{3 \, e^{2 \varphi } \, \chi }{(e^{\varphi} \, \chi -i )} \\[5mm]
 \dfrac{3 \, e^{2 \varphi } \, \chi }{(e^{\varphi } \, \chi -i)} & 3 \,  ( e^{\varphi } \, \chi^2+e^{-\varphi})
\end{array}
\right) \ .
\end{equation}

The two abelian isometries of the quaternionic metric (\ref{dsQK_nv=1&nh=1}) that are gauged correspond to Killing vectors of the form
\begin{equation}
\label{Killing_vectors_nv=1&nh=1}
\begin{array}{lll}
k_{\mathbb{R}} &=& \partial_{\sigma} \ , \\[2mm]
k_{\mathbb{U}} &=& - 3 \, ( \tilde{\zeta} \, \partial_{\zeta} - \zeta \, \partial_{\tilde{\zeta}} ) \ ,
\end{array}
\end{equation}
and have associated moment maps given by
\begin{equation}
\label{Momentum_maps_nv=1&nh=1}
\begin{array}{lll}
P^{+}_{\mathbb{R}} = 0
&\hspace{8mm} , \hspace{8mm} &
P^{3}_{\mathbb{R}} = - \tfrac{1}{2} e^{2 \phi} \ , \\[2mm]
P^{+}_{\mathbb{U}} = 3 \, e^{\phi} \, (\tilde{\zeta} - i \, \zeta)
&\hspace{8mm} , \hspace{8mm}&
P^{3}_{\mathbb{U}} =  3 \, \left( 1 - \frac{1}{4} \, e^{2 \phi} \,  (\zeta^2+\tilde{\zeta}^2)\right)  \ ,
\end{array}
\end{equation}
with $P^{+}\equiv P^{1}+ i P^{2}$. These isometries are gauged using the embedding tensor in (\ref{Theta_tensor_N2}) particularised to the case $\,n_{v}=1\,$. Plugging such an embedding tensor, together with the Killing vectors (\ref{Killing_vectors_nv=1&nh=1}), into the covariant derivatives (\ref{Dq_N2}) one finds
\begin{equation}
\label{Dq_nv=1&nh=1}
\begin{array}{c}
D \sigma = d \sigma - g \, \mathcal{A}^0 + m \, \tilde{\mathcal{A}}_0
\hspace{5mm} , \hspace{5mm}
D \zeta = d \zeta  + 3 \, g \, \mathcal{A}^1  \tilde{\zeta}  
\hspace{5mm} , \hspace{5mm}
D \tilde{\zeta} = d \tilde{\zeta} - 3 \, g  \, \mathcal{A}^1 \zeta     \ .
\end{array}
\end{equation}

Using the above geometrical data, the algebraic set of attractor equations (\ref{attrac_eqs})-(\ref{extra_constraints}) was solved in full generality in \cite{Guarino:2017eag} for a symplectic vector of charges $\,{\mathcal{Q}^M}\,$ of the form (\ref{Q_vector_attractor}). The result is that two horizon configurations -- related to each other by a $\,\mathbb{Z}_{2}\,$ reflection of the charges -- exist with scalar fields and vector charges being fixed to
\begin{equation}
\label{horizon_conf_nv=1&nh=1_enhancement}
\zeta_{h} = \tilde{\zeta}_{h} = \sigma_{h}= 0 \ ,
\end{equation}
and
\begin{equation}
\label{horizon_conf_nv=1&nh=1}
\begin{array}{lcll}
\kappa \, m^{-1/6} \, g^{7/6} \, L_{\textrm{AdS}_{2}}= - \dfrac{1}{2\,\, 3^{1/4}} 
& , & 
\kappa \, m^{-1/3} \, g^{7/3} \, L^2_{\Sigma_{2}}= -\dfrac{1}{2\, \sqrt{3}}  & , \\[4mm]
m^{1/3} \, g^{-1/3} \,  e^{\phi_{h}} = \sqrt{2} 
& \hspace{5mm} , \hspace{5mm} & 
m^{-1/3} \, g^{1/3} \,  z_{h}=  e^{i \frac{\pi}{3}}   & ,  \\[4mm]
m^{-2/3} \, g^{5/3} \, \mathfrak{p}^{0} = \pm \, \dfrac{1}{6}
& , & 
m^{1/3} \, g^{2/3} \, \mathfrak{e}_{0} = \pm \, \dfrac{1}{6}  & , \\[3mm]
g \, p^{1}=\mp \, \dfrac{1}{3} 
& , & 
m^{-1/3} \, g^{4/3} \, e_{1} = \pm \, \dfrac{1}{2}  & .
\end{array}
\end{equation}
The horizon must be of hyperbolic type $\,(\kappa=-1)$ for $\,L_{\textrm{AdS}_{2}}>0\,$ and $\, L^2_{\Sigma_{2}}>0\,$, and the phase $\,\beta\,$ in (\ref{attrac_eqs}) gets fixed to $\,\beta=\frac{\pi}{3} \mp \frac{\pi}{2}\,$. The requirements $\,\zeta_{h} = \tilde{\zeta}_{h} = 0\,$ and $\,\sigma_{h}=0\,$ respectively follow from the constraint $\,\left\langle  \mathcal{K}^{u} , \mathcal{V} \right\rangle=0\,$ in (\ref{attrac_eqs}) and from (\ref{EOM_A0tilde}). A quick inspection of the covariant derivatives in (\ref{Dq_nv=1&nh=1}) shows that (\ref{horizon_conf_nv=1&nh=1_enhancement}) decouples the vector $\,\mathcal{A}^{1}\,$, equivalently $\,k_{\mathbb{U}}=0\,$, and produces a $\,\textrm{U}(1)_{\mathbb{U}}\,$ symmetry enhancement in the truncation.

\subsection{$\mathcal{M}_{\textrm{SK}}=\textrm{SU(1,1)}/\textrm{U}(1)\,$ and $\,\mathcal{M}_{\textrm{QK}}=\textrm{G}_{2(2)}/\textrm{SO}(4)\,$}
\label{sec:nv=1&nh=2}

The next model extends the setup in \cite{Guarino:2017eag} by adding an additional hypermultiplet and corresponds to a truncation preserving the smallest non-abelian subgroup $\,\textrm{G}_{0}=\textrm{SO}(3) \subset \textrm{ISO}(7)\,$. In this case, the Lagrangian (\ref{Lagrangian_N2}) describes $\,\mathcal{N}=2\,$ supergravity coupled to a vector multiplet $\,(n_{v}=1)\,$ and two hypermultiplets $\,(n_{h}=2)\,$ with 
\begin{equation}
\mathcal{M}_{\textrm{SK}}=\frac{\textrm{SU(1,1)}}{\textrm{U}(1)}  
\hspace{8mm} \textrm{ and } \hspace{8mm}
\mathcal{M}_{\textrm{QK}}=\frac{\textrm{G}_{2(2)}}{\textrm{SO}(4)} \ .
\end{equation}
Since the extension only involves the hypermultiplet sector, the holomorphic sections, prepotential, vector kinetic terms and generalised theta angles as well as the kinetic term for the complex scalar $\,z\,$ are still given by (\ref{Xsections_nv=1&nh=1}), (\ref{F_prepot_nv=1&nh=1}), (\ref{NMat_nv=1&nh=1}) and (\ref{dsSK_nv=1&nh=1}). 

Regarding the geometrical description of $\,{\mathcal{M}_{\textrm{QK}}=\textrm{G}_{2(2)}/\textrm{SO}(4)}$, we fetch results from \cite{Erbin:2014hsa} (see also \cite{Fre:2014pca}). Denoting coordinates in $\,\mathcal{M}_{\textrm{QK}}\,$ by $\,q^{u}\,=(\tilde{\varphi} \,,\,\tilde{\chi} \,,\,\phi \,,\, \sigma \,,\,\zeta^{A} \,,\,\tilde{\zeta}_{A})$, with $\,A=0,1\,$, the quaternionic metric is given by
\begin{equation}
\label{dsQK_nv=1&nh=2}
\hspace{-2mm}
\begin{array}{lll}
ds_{\textrm{QK}}^2 &=&  3 \, d \tilde{\varphi} \, d\tilde{\varphi} + \frac{3}{4} \, e^{4\tilde{\varphi}} \, d\tilde{\chi} \, d\tilde{\chi} 
+  \, d \phi \, d \phi  - \frac{1}{4} \, e^{2\phi} \,   (d \vec{\zeta}\,)^{T} \,  \mathbb{C}\mathbb{M}_{4} \, d\vec{\zeta}  \\[2mm]
	&+& \frac{1}{4} e^{4 \phi} \left[ d \sigma + \frac{1}{2}\, (\vec{\zeta}\,)^{T} \, \mathbb{C} \, d \vec{\zeta} \,  \right] \,  \left[ d \sigma + \frac{1}{2}\, (\vec{\zeta}\,)^{T} \, \mathbb{C} \, d \vec{\zeta} \, \right] \ .
\end{array}
\end{equation}
The scalar matrix $\,\mathbb{M}_{4}\,$ entering (\ref{dsQK_nv=1&nh=2}) depends on the complex scalar  $\,\tilde{z}^{1} \equiv \tilde{z} = \tilde{\chi} + i \, e^{-2 \tilde{\varphi}}\,$ parameterising the special K\"ahler submanifold $\,\mathcal{M}_{\widetilde{\textrm{SK}}}=\textrm{SU}(1,1)/\textrm{U}(1)\,$ employed to construct $\,\mathcal{M}_{\textrm{QK}}\,$ via the c-map, and reads
\begin{equation}
\mathbb{M}_{4} = e^{6 \tilde{\varphi}}
\left(
\begin{array}{cccc}
- \tilde{\chi}^3 & 3 \tilde{\chi}^2 & 1  &  \tilde{\chi} \\[2mm]
- \tilde{\chi}^2 \left(\tilde{\chi}^2+e^{-4 \tilde{\varphi}}\right) &  \tilde{\chi} \left(3 \tilde{\chi}^2+2 e^{-4 \tilde{\varphi}}  \right) & \tilde{\chi} & \tilde{\chi}^2+\frac{1}{3} e^{-4 \tilde{\varphi}}  \\[2mm]
- \left(\tilde{\chi}^2+e^{-4 \tilde{\varphi}}\right)^3 & 3  \tilde{\chi} \left( \tilde{\chi}^2+e^{-4 \tilde{\varphi}}\right)^2 & \tilde{\chi}^3 &  \tilde{\chi}^2 \left(\tilde{\chi}^2+e^{-4 \tilde{\varphi}} \right) \\[2mm]
3 \tilde{\chi} \left(\tilde{\chi}^2+e^{-4 \tilde{\varphi}}\right)^2 & -3  \left(3  \tilde{\chi}^4+4 e^{-4 \tilde{\varphi}} \tilde{\chi}^2+e^{-8 \tilde{\varphi}}\right) & -3  \tilde{\chi}^2 & - \tilde{\chi} \left(3  \tilde{\chi}^2+2 e^{-4 \tilde{\varphi}}\right) \\
\end{array}
\right) \ .
\end{equation}
The geometrical data of the submanifold $\,\mathcal{M}_{\widetilde{\textrm{SK}}}\,$ is specified in terms of holomorphic sections $\,\tilde{Z} = (1 \,,\, \tilde{z} \,,\, \tilde{z}^3 \,,\,-3 \tilde{z}^2)\,$ which are compatible with a prepotential of the form
\begin{equation}
\widetilde{\mathcal{F}} = - \frac{(\tilde{Z}^{1})^3}{\tilde{Z}^{0}} \ .
\end{equation}
The K\"ahler potential for $\,\mathcal{M}_{\widetilde{\textrm{SK}}}\,$ enters the moment maps of the $\,\textrm{U}(1)_{\mathbb{U}}\,$ isometry being gauged. It follows the standard definition in (\ref{K_pot}) and reads $\,{\widetilde{K}=-\log(i \, \langle \tilde{Z}, \bar{\tilde{Z}} \rangle)}\,$.

The two isometries of the quaternionic metric (\ref{dsQK_nv=1&nh=2}) that are gauged in this truncation are specified by Killing vectors
\begin{equation}
\label{Killing_vectors_nv=1&nh=2}
\begin{array}{lll}
k_{\mathbb{R}} &=& \partial_{\sigma} \ , \\[2mm]
k_{\mathbb{U}} &=& \left[ (\mathbb{U} \, \tilde{Z})^{A} \partial_{\tilde{Z}^{A}} + \textrm{c.c.}\right] + (\mathbb{U} \, \vec{\zeta} \, )^{T}  \, \partial_{\vec{\zeta}} \ ,
\end{array}
\end{equation}
where the $\,\mathbb{U}\,$ matrix is given by
\begin{equation}
\mathbb{U} = 
\left(
\begin{array}{rrrr}
 0 & 3 & 0 & 0 \\
 -1 & 0 & 0 & -\frac{2}{3} \\
 0 & 0 & 0 & 1 \\
 0 & 6 & -3 & 0 \\
\end{array}
\right) \ .
\end{equation}
Following the terminology in \cite{Erbin:2014hsa}, the isometries (\ref{Killing_vectors_nv=1&nh=2}) are identified with duality symmetries and have associated moment maps of the form
\begin{equation}
\begin{array}{lll}
P^{+}_{\mathbb{R}} = 0
&\hspace{8mm} , \hspace{8mm} &
P^{3}_{\mathbb{R}} = - \tfrac{1}{2} e^{2 \phi} \ ,\\[2mm]
P^{+}_{\mathbb{U}} = - \sqrt{2} \,  e^{\frac{\widetilde{K}}{2}+\phi}  \tilde{Z}^{T} \mathbb{C} \, \mathbb{U} \, \vec{\zeta} 
&\hspace{8mm} , \hspace{8mm}&
P^{3}_{\mathbb{U}} = -  \tfrac{1}{4} e^{2 \phi} (\vec{\zeta}\,)^{T} \mathbb{C} \, \mathbb{U} \, \vec{\zeta}  + e^{\widetilde{K}} \tilde{Z}^{T} \mathbb{C} \, \mathbb{U} \, \bar{\tilde{Z}} \ .
\end{array}
\end{equation}
The embedding tensor in this model is still given by (\ref{Theta_tensor_N2}) with $\,n_{v}=1\,$. After using the Killing vectors in (\ref{Killing_vectors_nv=1&nh=2}) one finds covariant derivatives
\begin{equation}
\label{Dq_nv=1&nh=2}
\begin{array}{c}
D \tilde{\varphi} = d \tilde{\varphi} -  g \, \mathcal{A}^1 \tilde{\chi} 
\hspace{3mm} , \hspace{3mm}
D \tilde{\chi}  = d \tilde{\chi} + g \, \mathcal{A}^1 (1-e^{-4 \tilde{\varphi}} + \tilde{\chi}^2) \hspace{3mm} , \hspace{3mm}
D \sigma = d \sigma - g \, \mathcal{A}^0 + m \, \tilde{\mathcal{A}}_0 \ , \\[3mm]
\,\,\,\,\,\,\,\,\,\,\,D \zeta^{0} = d \zeta^{0}  - 3 \, g \, \mathcal{A}^1  \zeta^{1}
\hspace{3mm} , \hspace{3mm}
D \zeta^{1} = d \zeta^{1} + g  \, \mathcal{A}^1 (\zeta^{0} + \frac{2}{3} \tilde{\zeta}_{1}) \,\,\ , \\[3mm]
\,\,\,\,\,\,\,\,\,\,\,\, D \tilde{\zeta}_{0} = d \tilde{\zeta}_{0} - g  \, \mathcal{A}^1 \tilde{\zeta}_{1}
\,\,\,\,\hspace{3mm} , \hspace{3mm}
D \tilde{\zeta}_{1} = d \tilde{\zeta}_{1} + 3 \, g  \, \mathcal{A}^1 (\tilde{\zeta}_{0} -2 \zeta^{1})       \ .
\end{array}
\end{equation}

Let us now move to analyse the attractor equations (\ref{attrac_eqs})-(\ref{quant_cond}) using the vector of charges $\,\mathcal{Q}^{M}\,$ in (\ref{Q_vector_attractor}) and the quaternionic geometrical data presented above. By looking at the last equation in (\ref{attrac_eqs}), which is independent of the vector of charges, one finds that
\begin{equation}
\label{zt=zetaA=0}
\tilde{z}_{h}= i
\hspace{10mm} \textrm{ and } \hspace{10mm} 
\zeta^{A}{}_{h} = \tilde{\zeta}_{A \, h} = 0 \ .
\end{equation}
This renders the model with $\,\mathcal{M}_{\widetilde{\textrm{SK}}}=\textrm{SU}(1,1)/\textrm{U}(1)\,$ equivalent to the one with trivial $\,\mathcal{M}_{\widetilde{\textrm{SK}}}\,$ in what regards the classification of BPS horizon configurations. As a result, the only solution to the attractor mechanism is the one in (\ref{horizon_conf_nv=1&nh=1}). From the covariant derivatives in (\ref{Dq_nv=1&nh=2}), one sees that (\ref{zt=zetaA=0}) decouples the vector $\,\mathcal{A}^{1}\,$, namely $\,k_{\mathbb{U}}=0\,$, and the case presented in the previous section is recovered.

\subsection{$\mathcal{M}_{\textrm{SK}}=[\textrm{SU(1,1)}/\textrm{U}(1)]^3\,$ and $\,\mathcal{M}_{\textrm{QK}}=\textrm{SU(2,1)}/(\textrm{SU}(2) \times \textrm{U}(1))\,$}
\label{sec:nv=3&nh=1}

The last model extends the setup of \cite{Guarino:2017eag} by adding extra matter multiplets in the form of two vector multiplets. It describes $\,\mathcal{N}=2\,$ supergravity coupled to three vector multiplets $\,(n_{v}=3)\,$ and the universal hypermultiplet $\,(n_{h}=1)\,$ and has
\begin{equation}
\mathcal{M}_{\textrm{SK}}= \left[ \,  \frac{\textrm{SU(1,1)}}{\textrm{U}(1)} \, \right]^3
\hspace{8mm} \textrm{ and } \hspace{8mm}
\mathcal{M}_{\textrm{QK}}=\frac{\textrm{SU}(2,1)}{\textrm{SU}(2) \times \textrm{U}(1)} \ .
\end{equation}

Despite the presence of three vector multiplets, only the combination $\,{\mathcal{A}_{\mathbb{U}} \equiv \mathcal{A}^{1}+\mathcal{A}^{2}+\mathcal{A}^{3}}\,$ is associated with the $\,\textrm{U}(1)_{\mathbb{U}}\,$ factor of the gauging, as it can be seen from the covariant derivatives in (\ref{Dq_N2_massive_IIA}). Therefore, none of the scalars in the universal hypermultiplet are charged under the vectors associated with the two orthogonal combinations of U(1)'s, which turn to consistently decouple. This model is the massive IIA analogue of the $\,\textrm{U}(1)^4 \subset \textrm{SO(8)}\,$ invariant \mbox{STU-model} from M-theory. However, as discussed in the introduction, in the massive IIA case one is forced to keep the universal hypermultiplet which contains the scalars $\,\sigma\,$ and $\,(\zeta\,,\,\tilde{\zeta})\,$ that are charged under the gauge group $\,\textrm{G}=\mathbb{R} \times \textrm{U}(1)_{\mathbb{U}}\,$. Therefore the model in this section corresponds to a truncation with $\,{\textrm{G}_{0}=\textrm{U}(1)^2 \subset \textrm{U}(1)^2 \times \mathbb{R} \times \textrm{U}(1)_{\mathbb{U}} \subset \textrm{ISO}(7)  }\,$.

The holomorphic sections describing the SK manifold are the non-isotropic generalisation of the ones in (\ref{Xsections_nv=1&nh=1}) and take the form
\begin{equation}
\label{Xsections_nv=3&nh=1}
(X^{0},X^{1},X^{2},X^{3},F_{0},F_{1},F_{2},F_{3}) =  (-z^{1} z^{2} z^{3}\,,\,-z^{1}\,,\,-z^{2}\,,\,-z^{3}\,,\,1\,,\, z^{2} z^{3}\,,\, z^{3} z^{1}\,,\, z^{1} z^{2}) \ ,
\end{equation}
which this time are consistent with the square-root prepotential
\begin{equation}
\label{F_prepot_nv=3&nh=1}
\mathcal{F} = -2 \sqrt{X^0 \, X^1 \, X^2 \, X^3} \ .
\end{equation}

Denoting $\,z^{i} = - \chi_{i} + i \, e^{-\varphi_{i}}\,$ the complex coordinates in the SK manifold, the metric (\ref{dsSK_N2}) takes the form
\begin{equation}
\label{dsSK_nv=3&nh=1}
ds_{\textrm{SK}}^2  = \frac{1}{4} \sum_{i}\frac{dz^{i} \, d\bar{z}^{\bar{i}}}{(\textrm{Im} z^{i})^2} \ ,
\end{equation}
and the QK metric is still given by the one of the universal hypermultiplet (\ref{dsQK_nv=1&nh=1}). The kinetic terms and the generalised theta angles for the vector fields are encoded in the matrix
\begin{equation}
\mathcal{N}_{\Lambda\Sigma}=
\frac{1}{n}
\left(
\begin{array}{cccc}
 -i e^{\varphi_{1}+\varphi_{2}+\varphi_{3}} & n_{1} & n_{2} & n_{3}\\
n_{1} & -i
   e^{\varphi_{1}-\varphi_{2}-\varphi_{3}} \, c_{2} \, c_{3} & 
   n_{12} & n_{13}  \\
n_{2} &
n_{12} & -i e^{-\varphi_{1}+\varphi_{2}-\varphi_{3}} \, c_{1} \, c_{3} & 
n_{23}\\
n_{3}  & n_{13} & n_{23} & -i e^{-\varphi_{1}-\varphi_{2}+\varphi_{3}} \, c_{1} \, c_{2}
\end{array}
\right) \ ,
\end{equation}
with $\,c_{i} \equiv ( 1+ e^{2 \varphi_{i}} \, \chi_{i}^2)\,$ and where, in order to shorten expressions, we have introduced the quantities
\begin{equation}
\begin{array}{llll}
n & \equiv & \Big( 1+ \displaystyle\sum_{k} e^{2 \varphi_{k}}  \chi_{k}^{2}  \Big)  + 2 \, i \, e^{\varphi_{1}+\varphi_{2}+\varphi_{3}}\,  \chi_{1} \, \chi_{2} \, \chi_{3} \ , &  \\[2mm]
n_{i} & \equiv & e^{2 \varphi_{i}} \chi_{i}+i \, e^{\varphi_{1}+\varphi_{2}+\varphi_{3}} \chi_{j} \, \chi_{k} & \hspace{10mm} (i \neq j \neq k) \ ,\\[2mm]
n_{ij} & \equiv & e^{-\varphi_{k}} \, c_{k} \,  (e^{\varphi_{k}} \,  \chi_{k}+ i \,  e^{\varphi_{i} + \varphi_{j}} \, \chi_{i} \chi_{j} )  & \hspace{10mm} (i \neq j \neq k) \ .
\end{array}
\end{equation}

As in the model with one vector multiplet, the two isometries of $\,\mathcal{M}_{\textrm{QK}}\,$ that are gauged correspond to the Killing vectors and moment maps in (\ref{Killing_vectors_nv=1&nh=1}) and (\ref{Momentum_maps_nv=1&nh=1}).\footnote{The overall factor of $\,3\,$ in $\,k_{\mathbb{U}}\,$, $\,P^{+}_{\mathbb{U}}\,$ and $\,P^{3}_{\mathbb{U}}\,$ must now be removed due to the non-isotropic ($n_{v}=3$) setup.} These isometries are again gauged using the dyonic embedding tensor in (\ref{Theta_tensor_N2}) particularised this time to the case $\,n_{v}=3\,$. The resulting covariant derivatives for the charged scalars in the universal hypermultiplet read
\begin{equation}
\label{Dq_nv=3&nh=1}
\begin{array}{c}
D \sigma = d \sigma - g \, \mathcal{A}^0 + m \, \tilde{\mathcal{A}}_0
\hspace{5mm} , \hspace{5mm}
D \zeta = d \zeta  + g \, \mathcal{A}_{\mathbb{U}} \,  \tilde{\zeta}  
\hspace{5mm} , \hspace{5mm}
D \tilde{\zeta} = d \tilde{\zeta} - g  \,  \mathcal{A}_{\mathbb{U}} \,  \zeta     \ ,
\end{array}
\end{equation}
with $\,\mathcal{A}_{\mathbb{U}} \equiv \mathcal{A}^1+\mathcal{A}^2+\mathcal{A}^3\,$. Note that this model reduces to the one in section~\ref{sec:nv=1&nh=1} if the three vector multiplets are identified. However, the presence of two linear combinations of abelian vectors not being coupled to the universal hypermultiplet becomes crucial to obtain a four-parameter family of BPS horizon configurations generalising the one in (\ref{horizon_conf_nv=1&nh=1}).

In the following we are solving the attractor equations (\ref{attrac_eqs})-(\ref{quant_cond}) using the symplectic vector of charges in (\ref{Q_vector_attractor}) with $\,n_{v}=3\,$. As in the previous examples, the last equation in (\ref{attrac_eqs}) and the duality relation (\ref{EOM_A0tilde}) require
\begin{equation}
\label{symmetry_enhancement_nv=3}
\zeta_{h}=\tilde{\zeta}_{h}=0
\hspace{5mm} \textrm{ and } \hspace{5mm} 
\sigma_{h}=0 \ ,
\end{equation}
thus decoupling the vector $\,\mathcal{A}_{\mathbb{U}}\,$ in (\ref{Dq_nv=3&nh=1}) and producing a $\,\textrm{U}(1)_{\mathbb{U}}\,$ symmetry enhancement in the truncation. In addition, the last equation in (\ref{attrac_eqs}) imposes a constraint on the scalars $\,z_{h}^{i}\,$ in the vector multiplets of the form
\begin{equation}
\label{SK_attractor}
\prod_{i} \, z_{h}^{i} = -\frac{m}{g} \ ,
\end{equation}
which allows us to express one of the fields, let us say $\,z_{h}^{k}\,$, in terms of the others as
\begin{equation}
\label{SK_attractor_sol}
z_{h}^{k}=- \frac{m}{g} \, \frac{1}{z_{h}^{i} z_{h}^{j}}
\hspace{5mm} \textrm{ with } \hspace{5mm} i \neq j \neq k \ . 
\end{equation}
The relation (\ref{SK_attractor_sol}) requires non-vanishing axions for $\,\textrm{Im}z_{h}^{k} \neq 0\,$. Note also that plugging (\ref{SK_attractor_sol}) into (\ref{F_prepot_nv=3&nh=1}) gives $\,{\mathcal{F}(z_{h}^{i},z_{h}^{j})=-2 \, (m/g)}\,$.

The two (complex) scalars $\,(z_{h}^{i},z_{h}^{j})\,$ that remain unfixed in (\ref{SK_attractor_sol}) yield a (real) four-parameter family of BPS horizon configurations. Moreover, the first condition in (\ref{extra_constraints}) and the quantisation condition (\ref{quant_cond}) require
\begin{equation}
\label{extra_cond_STU_model}
\mathfrak{e}_{0} \, m - g \, \mathfrak{p}^{0} = 0 
\hspace{10mm} \textrm{ and } \hspace{10mm} 
\displaystyle \sum_{i} p^{i} = \mp \, \frac{1}{g} \ .
\end{equation} 
After imposing (\ref{extra_cond_STU_model}), the first and second equations in (\ref{attrac_eqs}) completely determine the rest of the quantities at the horizon:
\begin{equation}
\label{Attractor_config}
e^{2 \phi_{h}(z_{h}^{i},z_{h}^{j})}
\hspace{3mm} , \hspace{3mm} 
L^2_{\textrm{AdS}_{2}}(z_{h}^{i},z_{h}^{j})
\hspace{3mm} , \hspace{3mm} 
\kappa \, L^{2}_{\Sigma_{2}}(z_{h}^{i},z_{h}^{j})
\hspace{3mm} \textrm{ and } \hspace{3mm} 
\mathcal{Q}(z_{h}^{i},z_{h}^{j}) \ ,
\end{equation}
and the phase $\,\beta\,$ in (\ref{EqZ+L=0}). They are expressed in terms of the two complex scalars $\,(z_{h}^{i},z_{h}^{j})\,$ in (\ref{SK_attractor_sol}) and the gauging parameters $\,(g,m)\,$. Note that, instead of trying to invert the relations $\,\mathcal{Q}(z_{h}^{i},z_{h}^{j})\,$ to parameterise the space of BPS horizon configurations in terms of vector charges, we prefer to use the values of the complex scalars. This will be more convenient later on when exploring the region of the parameter space giving rise to physically acceptable horizons. Lastly, we have verified that the horizon configurations (\ref{Attractor_config}) extremise the effective black hole potential (\ref{Veff}).

The explicit expressions for the functions in (\ref{Attractor_config}), especially for the charges, are not very enlightening at this stage. The value of the dilaton field in the universal  hypermultiplet at the horizon reads
\begin{equation}
\label{Phi_att}
e^{2 \phi_{h}} = \frac{1}{N(z_{h}^{i},z_{h}^{j})} \left[ \, \textrm{Im}z_{h}^{i} \, \textrm{Im}z_{h}^{j} + \frac{g}{m} \, \textrm{Im}(z_{h}^{i} \, z_{h}^{j}) \, \big(  z^{(2,0,1,0)} + z^{(1,0,2,0)} + z^{(0,2,1,0)} +  z^{(1,0,0,2)} \big) \, \right] \ ,
\end{equation}
where we have introduced the short-hand notation
\begin{equation}
z^{(n_{1},n_{2},n_{3},n_{4})} \equiv (\textrm{Im}z_{h}^{i})^{n_{1}} (\textrm{Re}z_{h}^{i})^{n_{2}} (\textrm{Im}z_{h}^{j})^{n_{3}} (\textrm{Re}z_{h}^{j})^{n_{4}} \ ,
\end{equation}
and the function
\begin{equation}
\label{N_function}
N(z_{h}^{i},z_{h}^{j}) = z^{(2,0,2,0)} + z^{(1,1,1,1)} + z^{(2,0,0,2)} +  z^{(0,2,2,0)} \ .
\end{equation}
The radius of the AdS$_{2}$ factor of the metric at the horizon reads
\begin{equation}
\label{LAdS_att}
L^2_{\textrm{AdS}_{2}} =  \frac{2 m}{g} \, \frac{\textrm{Im}z_{h}^{i} \, \textrm{Im}z_{h}^{j} \, \textrm{Im}(z_{h}^{i} \, z_{h}^{j})    }{ m^2 - 2  g m \, \textrm{Re}\big( \, z_{h}^{i} \, z_{h}^{j} \, (z_{h}^{i}+z_{h}^{j}) \, \big)  + g^2  \, |z_{h}^{i} \, z_{h}^{j} \, (z_{h}^{i}+z_{h}^{j})|^2} \ .
\end{equation}
From the vanishing of (\ref{EqZ+L=0}), the radius of the $\,\Sigma_{2}\,$ factor can be written as
\begin{equation}
\label{LSigma2_att}
\kappa \, L^{2}_{\Sigma_{2}} =  i \, \frac{\mathcal{Z}(z_{h}^{i},z_{h}^{j})}{\mathcal{L}(z_{h}^{i},z_{h}^{j})}
\hspace{5mm}
\textrm{ with } 
\hspace{5mm}
i \neq j  \ ,
\end{equation}
in terms of the superpotential $\,\mathcal{L}\,$ and the central charge $\,\mathcal{Z}\,$ evaluated at the solution (\ref{Attractor_config}) of the attractor equations. The superpotential is given by
\begin{equation}
\label{L_att}
\mathcal{L}(z_{h}^{i},z_{h}^{j})  = \pm \, e^{\frac{K}{2}}  \, \Big[ \, \frac{m}{z_{h}^{i} \, z_{h}^{j}} - g \, (z_{h}^{i} + z_{h}^{j}) \, \Big]
\hspace{5mm} \textrm{ with } \hspace{5mm}
e^{K}  = \frac{g}{8 m} \, \frac{ |z_{h}^{i} \, z_{h}^{j}|^2}{\textrm{Im}z_{h}^{i} \, \textrm{Im}z_{h}^{j} \, \textrm{Im}(z_{h}^{i}z_{h}^{j})} \ .
\end{equation}
The central charge is given by
\begin{equation}
\label{Z_att}
\mathcal{Z}(z_{h}^{i},z_{h}^{j}) = \mp \, \frac{i}{4} \, e^{-\frac{K}{2}}   \, |z_{h}^{i} \, z_{h}^{j}|^2 \, \Big[ \, \frac{m}{z_{h}^{i} z_{h}^{j}} - g \,  (z_{h}^{i}+z_{h}^{j}) \, \Big] \, \frac{N(z_{h}^{i},z_{h}^{j})}{D(z_{h}^{i},z_{h}^{j})} \ ,
\end{equation}
where we have introduced the function
\begin{equation}
\label{D_function}
\begin{array}{lll}
D(z_{h}^{i},z_{h}^{j}) & = &  m^2 \,  \big( \, z^{(2,0,0,2)} +  \frac{1}{2} \,  z^{(1,1,1,1)} \, \big)  \\[2mm]
&-& 2  g m \,\,\,  \big( \,\,\, z^{(4,0,0,3)}  + 2  \,   z^{(4,0,2,1)} + \, z^{(3,1,3,0)}    +   z^{(3,1,1,2)}   +   z^{(3,0,1,3)}    \\[1mm]
&&  \,\,\,\,\,\,\,\,\,\,\,\,\,\,  + \,\,  z^{(2,2,0,3)}  + 3  \,  z^{(2,2,2,1)}  + \,   z^{(2,1,0,4)}  + \,  z^{(1,3,1,2)}   \, \big) \\[2mm]
& + & g^{2} \,\, \big(   \, \phantom{+}  z^{(6,0,0,4)}  +  \phantom{0} \,  z^{(5,1,1,3)}  + \phantom{0} \,  z^{(4,0,2,4)}   +  \phantom{0} \, z^{(2,4,0,4)}  +  \phantom{0} \,  z^{(2,2,0,6)} + \phantom{0} \,   z^{(6,0,2,2)}         \\[1mm]
& &\,\,\,\,\,\,\,  +  \,\, \phantom{0}   z^{(6,0,4,0)} +  \phantom{0}\,  z^{(1,5,1,3)}   -  \phantom{0} \,  z^{(5,1,3,1)} - \phantom{0} \,  z^{(3,1,1,5)}      -  \phantom{0} \,  z^{(5,0,5,0)}  + \phantom{0} \,  z^{(1,4,1,4)}    \\[1mm]
& & \,\,\,\,\,\,\,    + \,  2 \,  z^{(4,2,4,0)}  -  2 \, z^{(3,3,3,1)}   + 2 \,  z^{(4,2,0,4)}   + 2 \,  z^{(4,1,0,5)}      - 2 \, z^{(4,1,2,3)} + 2 \,  z^{(3,3,1,3)} \\[1mm]
& &  \,\,\,\,\,\,\,   + \,  2 \,  z^{(2,3,0,5)}    - 2 \, z^{(3,2,1,4)}   - 2 \, z^{(4,1,4,1)}   - 4 \, z^{(3,2,3,2)}   - 4 \,  z^{(5,0,1,4)}   - 6 \, z^{(5,0,3,2)}   \\[1mm]
& & \,\,\,\,\,\,\,  + \, 3  \,  z^{(4,2,2,2)}   + 3  \,  z^{(2,4,2,2)}  \, \big)  \quad  + \quad  \textrm{ perm } \ .
\end{array}
\end{equation}
The permutation (perm) terms in (\ref{D_function}) account for the exchange $\,z_{h}^{i} \leftrightarrow z_{h}^{j}\,$ and correspond to terms with $\,(n_{1},n_{2})  \leftrightarrow (n_{3},n_{4}) \,$. Finally, the expressions for the charges associated with the graviphoton and its magnetic dual in (\ref{Q_vector_attractor}) read
\begin{eqnarray}
\label{p0e0_att}
\mathfrak{p}^{0} &=& \dfrac{m}{g} \, \mathfrak{e}_{0} \ , \\
\mathfrak{e}_{0} &=& \pm \, \Big[ \, g \, \textrm{Re}(z_{h}^{i} z_{h}^{j}) \, (z^{(2,0,0,1)}+z^{(0,1,2,0)}+z^{(0,2,0,1)}+z^{(0,1,0,2)})  -  m \, \textrm{Re}z_{h}^{i} \, \textrm{Re}z_{h}^{j}  \, \Big] \, \dfrac{N(z_{h}^{i} , z_{h}^{j})}{D(z_{h}^{i} , z_{h}^{j})} \nonumber \ .
\end{eqnarray}
Similar expressions are found for the charges $\,(p^{k},e_{k})\,$ in terms of the scalars at the horizon, although we are not displaying them here\footnote{Like the charges $\,(\mathfrak{p}^{0},\mathfrak{e}_{0})\,$ in (\ref{p0e0_att}), and also $\,L^{2}_{\Sigma_{2}}\,$ in (\ref{LSigma2_att}), the charges $\,(p^{k},e_{k})\,$ are proportional to $\,D(z_{h}^{i} , z_{h}^{j})^{-1}\,$. Therefore, all these quantities blow-up whenever $\,D(z_{h}^{i} , z_{h}^{j})\,$ vanishes.}. Note that the quantities (\ref{Phi_att}), (\ref{LAdS_att}), (\ref{LSigma2_att}) and (\ref{p0e0_att}) are consistently symmetric under the exchange $\,z_{h}^{i} \leftrightarrow z_{h}^{j}\,$.

An important quantity that can be computed solely from the horizon data is the gravitational entropy density. At leading order, it is given by the Bekenstein--Hawking formula
\begin{equation}
\label{entropyBH}
s = \frac{V_{\textrm{eff}}(z_{h}^{i},z_{h}^{j}) }{4} = \frac{L^{2}_{\Sigma_{2}}(z_{h}^{i},z_{h}^{j}) }{4} = \frac{m}{2 \, \kappa \, g}  \, \textrm{Im}z_{h}^{i} \, \textrm{Im}z_{h}^{j} \, \textrm{Im}(z_{h}^{i}z_{h}^{j}) \, \frac{N(z_{h}^{i},z_{h}^{j})}{D(z_{h}^{i},z_{h}^{j})}  \ ,
\end{equation}
with $\,i \neq j\,$. The above entropy density may be relevant in massive IIA holography in light of the recent advances in black hole microstate counting in the STU-model from M-theory featuring FI gaugings \cite{Benini:2015eyy,Benini:2016rke}. Note however that, in the massive IIA setup, the non-compact gauging is associated with isometries of the universal hypermultiplet (FI terms in the moment maps are permitted only when there are no physical hypermultiplets).

\subsubsection*{Example: one-parameter families of hyperbolic/spherical horizons}

In order to assess the existence of new BPS horizon configurations, we start from the isotropic configuration in (\ref{horizon_conf_nv=1&nh=1}) and parametrically deviate from it by setting
\begin{equation}
\label{deformation_eps}
m^{-1/3} \, g^{1/3}  \,\, z_{h}^{1} = e^{i \frac{\pi}{3}} + \epsilon 
\hspace{6mm} \textrm{ , } \hspace{6mm}
m^{-1/3} \, g^{1/3}  \,\, z_{h}^{2} =  e^{i \frac{\pi}{3}} - \lambda \,  \epsilon \ ,
\end{equation}
in terms of a continuous deformation parameter $\,\epsilon\,$ and a sign $\,\lambda=\pm\,$. For the sake of definiteness, we have set $\,(z_{h}^i,z_{h}^j)=(z_{h}^1,z_{h}^2)\,$ and $\,z_{h}^k=z_{h}^3\,$ without loss of generality. For each choice of $\,\lambda\,$, (\ref{deformation_eps}) specifies a one-parameter slice within the four-parameter space of BPS horizon configurations previously obtained. The solution (\ref{horizon_conf_nv=1&nh=1}) corresponds to $\,\epsilon=0\,$, identifies $\,z_{h}^{1,2,3}=z_{h}=(m/g)^{\frac{1}{3}} \, e^{i \frac{\pi}{3}}\,$ and requires a hyperbolic horizon ($\kappa=-1$).

\begin{figure}[t!]
\begin{center}
\includegraphics[width=130mm,height=130mm,keepaspectratio]{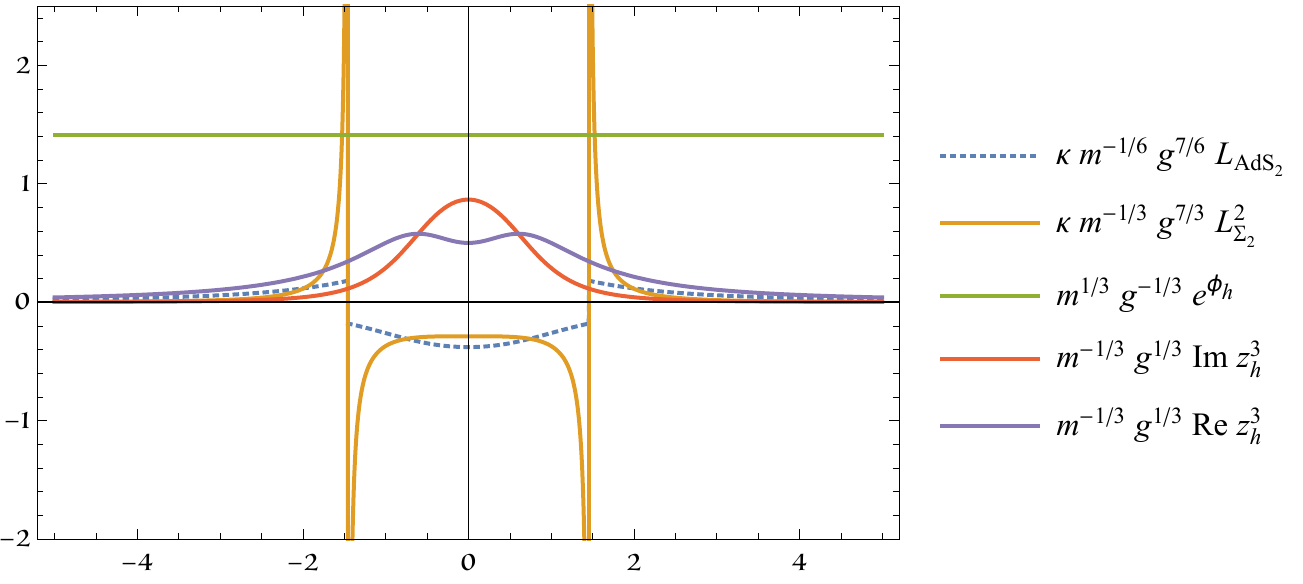}
\put(-240,-8){$\epsilon$ }
\caption{Horizon configurations as a function of the deformation parameter $\,\epsilon\,$ when $\,\lambda=+$. There is a transition from hyperbolic to spherical horizon at $\,|\epsilon_{\textrm{crit}}| = 3^{1/2} \, 2^{-1/4}$.
\label{Fig:epsilon_plot_lambda1}}
\end{center}
\end{figure}

Setting $\,\lambda=+\,$ the relevant quantities at the horizon are given by
\begin{equation}
m^{-1/3} \, g^{7/3}  \,\, L^2_{\textrm{AdS}_{2}} = \dfrac{3 \sqrt{3}}{4 \, (4 \,  \epsilon^4+6 \, \epsilon^2 + 9 )}
\hspace{5mm} , \hspace{5mm}
\kappa \, m^{-1/3} \, g^{7/3}  \,\, L^2_{\Sigma_{2}} = \dfrac{3 \sqrt{3}}{4 \, \epsilon^4-18}  \ ,
\end{equation}
and
\begin{equation}
m^{1/3} \, g^{-1/3}  \,\, e^{\phi_{h}} = \sqrt{2}
\hspace{6mm} , \hspace{6mm}
m^{-1/3} \, g^{1/3}  \, z_{h}^{3} = - \left(  e^{i \frac{2 \pi }{3}} - \epsilon^2  \right)^{-1}   \ .
\end{equation}
At the critical values $\,|\epsilon_{\textrm{crit}}| = 3^{1/2} \, 2^{-1/4}\,$ the radius $\,L^2_{\Sigma_{2}}\,$ and the vector of charges $\,\mathcal{Q}\,$ become singular due to the vanishing of $\,D(z_{h}^{1},z_{h}^{2})\,$ (see Figure~\ref{Fig:epsilon_plot_lambda1}). For $\,{|\epsilon|  < |\epsilon_{\textrm{crit}}| }\,$ the horizon is of hyperbolic type, whereas for $\,|\epsilon|  >  |\epsilon_{\textrm{crit}}|  \,$ the horizon is spherical. Note that the deformation does not affect the dilaton in the universal hypermultiplet which is still fixed at the horizon to the value in (\ref{horizon_conf_nv=1&nh=1}).

\begin{figure}[t!]
\begin{center}
\includegraphics[width=130mm,height=130mm,keepaspectratio]{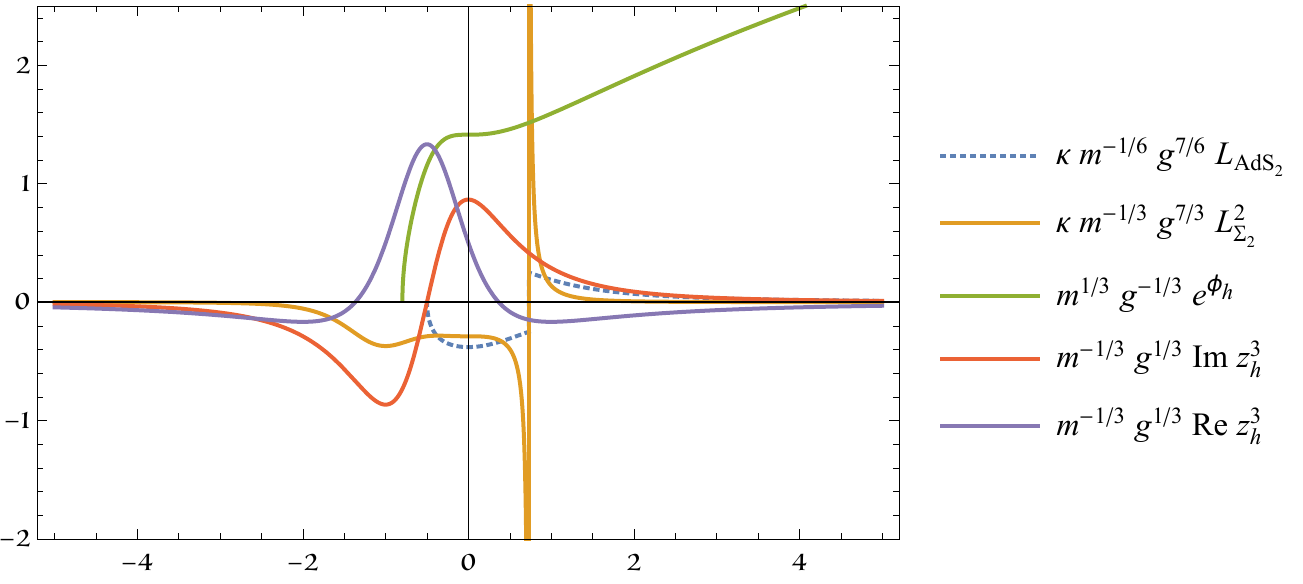}
\put(-240,-8){$\epsilon$ }
\caption{Horizon configurations as a function of the deformation parameter $\,\epsilon\,$ when $\,\lambda=-$. There is a transition from hyperbolic to spherical horizon at $\,\epsilon_{\textrm{crit}} \approx 0.7304$.
\label{Fig:epsilon_plot_lambda-1}}
\end{center}
\end{figure}

The situation changes when setting $\,\lambda=-\,$ as shown in Figure~\ref{Fig:epsilon_plot_lambda-1}. In this case, the value of the dilaton in the universal hypermultiplet at the horizon varies with the parameter $\,\epsilon\,$. The relevant quantities at the horizon are given by
\begin{equation}
\begin{array}{rll}
m^{-1/3} \, g^{7/3}  \,\, L^2_{\textrm{AdS}_{2}} &=& \dfrac{3 \sqrt{3} (2 \epsilon +1)}{16 \epsilon ^6+48 \epsilon ^5+96 \epsilon ^4+96 \epsilon ^3+72 \epsilon ^2+72 \epsilon +36} \ , \\[6mm]
\kappa \, m^{-1/3} \, g^{7/3}  \,\, L^2_{\Sigma_{2}} &=& \dfrac{3 \sqrt{3} \left(2 \epsilon ^2+2 \epsilon +1\right)}{16 \epsilon ^7+56 \epsilon ^6+96 \epsilon ^5+84 \epsilon ^4+24 \epsilon ^3-36 \epsilon^2-36 \epsilon -18}  \ , 
\end{array}
\end{equation}
and
\begin{equation}
\label{horizon_epsilon_3}
\begin{array}{lll}
m^{1/3} \, g^{-1/3}  \,\, e^{\phi_{h}} &=& \left[ \dfrac{2 \left(4 \epsilon ^3+6 \epsilon ^2+6 \epsilon +3\right)}{3 \left(2 \epsilon ^2+2 \epsilon +1\right)} \right]^{1/2} \ , \\[6mm]
m^{-1/3} \, g^{1/3}  \, z_{h}^{3} &=& - \left(  e^{i \frac{\pi }{3}} + \epsilon  \right)^{-2}   \ ,
\end{array}
\end{equation}
so the parameter $\,\epsilon\,$ turns to be bounded from below $\,\epsilon > -\frac{1}{2} \,$ in order to give acceptable values $\,L_{\textrm{AdS}_{2}} > 0\,$ and $\,\textrm{Im}z_{h}^{3} > 0\,$. The horizon is hyperbolic if $\, -\frac{1}{2} < \epsilon <  \epsilon_{\textrm{crit}}\,$ and spherical if $\,\epsilon >  \epsilon_{\textrm{crit}}\,$ with $\, \epsilon_{\textrm{crit}} \approx 0.7304 \,$. At the critical value, the radius $\,L^2_{\Sigma_{2}}\,$ and the vector of charges $\,\mathcal{Q}\,$ blow up due to the vanishing of $\,D(z_{h}^{1},z_{h}^{2})\,$.

\section{Conclusions}
\label{sec:conclus}

In this note we have investigated the attractor equations for static BPS black holes in various four-dimensional $\,\mathcal{N}=2\,$ gauged supergravities arising from the reduction of massive IIA supergravity on a six-sphere. The gauge group is $\,\textrm{G}=\mathbb{R} \times \textrm{U}(1)_{\mathbb{U}}\,$ and originates from the dyonic gauging of abelian isometries of the hypermultiplet moduli space. We have generalised the results in \cite{Guarino:2017eag} for the canonical model with one vector multiplet and the universal hypermultiplet to include extra matter both in the hypermultiplet and vector multiplet sectors.

The minimal extension of the hypermultiplet sector, namely having two hypermultiplets in the image of a c-map, does not allow for new BPS horizon configurations apart from the unique hyperbolic horizon found in \cite{Guarino:2017eag}. This follows from the general consideration that only the complex scalar in $\,\mathcal{M}_{\textrm{SK}}\,$ and the universal dilaton in $\,\mathcal{M}_{\textrm{QK}}\,$ can acquire non-trivial values $\,z_{h}\,$ and $\,e^{\phi_{h}}\,$ at the horizon by virtue of the attractor equations, which set $\,\tilde{z}_{h}=i\,$ and $\,\zeta^{A}{}_{h}=\tilde{\zeta}_{A \, h}=0\,$. As a consequence of the covariant derivatives in (\ref{Dq_nv=1&nh=2}), there is a $\,\textrm{U}(1)_{\mathbb{U}}\,$ symmetry enhancement in the truncation as none of the non-trivial scalars are charged under the vector in the vector multiplet. The relevant dynamics in this type of extensions of the hypermultiplet sector based on the c-map is then captured by the simplest model with only the universal hypermultiplet. 

The extension of the vector sector turns to be compatible with a richer set of horizon configurations including continuous parameters. In this note we have investigated the model with three vector multiplets and the universal hypermultiplet which is the massive IIA analogue of the STU-model from M-theory. The attractor equations can be solved in full generality giving rise to BPS horizon configurations that involve non-trivial values for the scalars in the vector multiplets $\,z_{h}^{i}\,$ and the dilaton in the universal hypermultiplet~$\,e^{\phi_{h}}$. The $\,\textrm{U}(1)_{\mathbb{U}}\,$ symmetry enhancement also occurs in this model as $\,{\zeta_{h}=\tilde{\zeta}_{h}=0}\,$. The horizons turn to depend on four continuous parameters as well as on the gauging parameters $\,(g,m)$. Dependending on the point in parameter space, they can have hyperbolic or spherical topology, thus generalising the results in \cite{Guarino:2017eag}.

Finally, in the model with three vector multiplets and the universal hypermultiplet, the gravitational entropy density associated with the horizons can be expressed in terms of the four continuous parameters allowed by the attractor equations and the gauging parameters $\,(g,m)$. In this note we found convenient to characterise the horizon configurations in terms of the values of the scalars $\,(z_{h}^{i},z_{h}^{j})\,$ rather than in terms of the charges $\,\mathcal{Q}\,$. A characterisation in terms of the latter requires the inversion of the non-linear algebraic relations $\,\mathcal{Q}(z_{h}^{i},z_{h}^{j})\,$ (see \textit{e.g.} (\ref{p0e0_att})) which are not straightforward to invert. For this reason, the entropy density in (\ref{entropyBH}) is not yet in a suggestive form to be used in massive IIA holography along the lines of the recent advances in the STU-model from M-theory featuring FI gaugings \cite{Benini:2015eyy,Benini:2016rke,Cabo-Bizet:2017jsl}. To make progress in this direction it is essential to carry out a study of static BPS black holes potentially flowing to the $\,\textrm{AdS}_{2} \times \Sigma_{2}\,$ horizon configurations discussed in this note. Altogether, it is important to get a better understanding of the massive~IIA on S$^{6}$/SYM-CS duality \cite{Guarino:2015jca,Schwarz:2004yj} beyond anti-de Sitter backgrounds. The less supersymmetric and hybrid SYM-CS nature of the duality makes it an interesting avenue to explore \cite{Fluder:2015eoa,Araujo:2016jlx,Araujo:2017hvi}. We hope to come back to some of these issues in the future.

\section*{Acknowledgements}

We are grateful to Nikolay Bobev for conversations and especially to Javier Tarr\'io for discussions and collaboration in related work. The work of AG is partially supported by a Marina Solvay fellowship and by F.R.S.-FNRS through the conventions PDRT.1025.14 and IISN-4.4503.15.

\bibliography{references}

\end{document}